\documentclass[aps,prb,twocolumn,9pt,superscriptaddress,citeautoscript,showkeys]{revtex4-2}

\pdfoutput=1
\usepackage{graphicx}
\usepackage{amsmath, amsfonts}
\usepackage{amssymb}
\usepackage[colorlinks=true,linkcolor=blue,citecolor=blue]{hyperref}
\usepackage[utf8]{inputenc}
\usepackage{float}

\usepackage{subcaption}
\usepackage{ragged2e}

\usepackage{etoolbox}
\patchcmd{\subsection}
  {\centering}
  {\raggedright}
  {}
  {}

\DeclareCaptionJustification{justified}{\justifying}
\setcitestyle{super}

\begin{document}

\title{Memristor-Driven Spike Encoding for Fully Implantable Cochlear Implants}


\author{Tímea Nóra Török}
\affiliation{Department of Physics, Institute of Physics, Budapest University of Technology and Economics, M\H{u}egyetem rkp. 3., H-1111 Budapest, Hungary.}
\affiliation{Institute of Technical Physics and Materials Science, HUN-REN Centre for Energy Research, Konkoly-Thege M. \'{u}t 29-33, 1121 Budapest, Hungary.}

\author{Roland Kövecs}
\affiliation{Department of Physics, Institute of Physics, Budapest University of Technology and Economics, M\H{u}egyetem rkp. 3., H-1111 Budapest, Hungary.}

\author{Ferenc Braun}
\affiliation{Institute of Technical Physics and Materials Science, HUN-REN Centre for Energy Research, Konkoly-Thege M. \'{u}t 29-33, 1121 Budapest, Hungary.}

\author{Zsigmond Pollner}
\affiliation{Department of Physics, Institute of Physics, Budapest University of Technology and Economics, M\H{u}egyetem rkp. 3., H-1111 Budapest, Hungary.}
\affiliation{HUN-REN–BME Condensed Matter Research Group,
Budapest University of Technology and Economics,
Műegyetem rkp. 3., H-1111 Budapest, Hungary.}

\author{Tamás Zeffer}
\affiliation{Institute of Technical Physics and Materials Science, HUN-REN Centre for Energy Research, Konkoly-Thege M. \'{u}t 29-33, 1121 Budapest, Hungary.}
\affiliation{Doctoral School on Material Sciences \& Technologies, Óbuda University, Bécsi str. 96/b, 1034 Budapest, Hungary.}

\author{Nguyen Quoc Khánh}
\affiliation{Institute of Technical Physics and Materials Science, HUN-REN Centre for Energy Research, Konkoly-Thege M. \'{u}t 29-33, 1121 Budapest, Hungary.}

\author{László Pósa}
\affiliation{Department of Physics, Institute of Physics, Budapest University of Technology and Economics, M\H{u}egyetem rkp. 3., H-1111 Budapest, Hungary.}
\affiliation{Institute of Technical Physics and Materials Science, HUN-REN Centre for Energy Research, Konkoly-Thege M. \'{u}t 29-33, 1121 Budapest, Hungary.}

\author{Péter Révész}
\affiliation{Department of Otorhinolaryngology--Head and Neck Surgery, Clinical Center, University of Pécs, Munkácsy M. u. 2, 7621 Pécs, Hungary.}

\author{Heungsoo Kim}
\affiliation{Naval Research Laboratory, 4555 Overlook Ave, Washington, DC 20375, USA.}

\author{Alberto Piqué}
\affiliation{Naval Research Laboratory, 4555 Overlook Ave, Washington, DC 20375, USA.}

\author{András Halbritter}
\affiliation{Department of Physics, Institute of Physics, Budapest University of Technology and Economics, M\H{u}egyetem rkp. 3., H-1111 Budapest, Hungary.}
\affiliation{HUN-REN–BME Condensed Matter Research Group,
Budapest University of Technology and Economics,
Műegyetem rkp. 3., H-1111 Budapest, Hungary.}

\author{János Volk}
\email{volk.janos@ek.hun-ren.hu}
\affiliation{Institute of Technical Physics and Materials Science, HUN-REN Centre for Energy Research, Konkoly-Thege M. \'{u}t 29-33, 1121 Budapest, Hungary.}

\begin{abstract}
{\emph{Objective}:
This work aims to demonstrate a low-power, biomimetic auditory sensing concept for fully implantable cochlear implants. The approach draws inspiration from the frequency selectivity and temporal encoding of the cochlea, and uses neuromorphic spike generation to replace conventional signal processing blocks. The goal is to establish a compact, energy-efficient front-end architecture suitable for future implantable systems.}

{\emph{Methods:}
An auditory sensing unit was implemented, consisting of a piezoelectric MEMS cantilever mechanically coupled to a single VO$_2$ nanogap Mott memristor-based oscillator. This configuration enables FFT-free, frequency-selective sensing and direct spike generation, forming a biomimetic auditory front end. The concept was experimentally examined using controlled mechanical excitation.}

{\emph{Results:}
The sensing unit exhibited frequency-selective detection of mechanical vibrations in the nanometer to tens-of-nanometers displacement range and generated biomimetic spiking waveforms. Spike rate-encoding of the input amplitude was demonstrated, with output spiking frequencies tunable between approximately 100 Hz and 1 kHz depending on the excitation level. The waveform was finally converted to a biphasic shape suitable for cochlear implant stimulation.}

{\emph{Significance:}
Temporal encoding is fundamental to natural auditory signal processing in the nervous system. By implementing this principle through neuromorphic spike encoding, the proposed approach can provide significant benefits for cochlear implants. In addition, the circuit has the potential to reduce footprint, energy consumption, and latencies compared with current commercial solutions.}
\end{abstract}


\date{\today}
\maketitle

\section{Introduction}
\label{sec:introduction}
Hearing loss affects millions of people worldwide, profoundly impacting communication, social interaction, and quality of life \cite{zhang_bidirectional_2024}. In 2019, over 403 million people had moderate to severe hearing loss, and this number is expected to nearly double by 2050 \cite{haile_hearing_2021}.
Cochlear implants (CIs) are medical devices designed to restore hearing in individuals with severe to profound sensorineural hearing loss by bypassing the damaged cochlea \cite{zeng_cochlear_2008}. They are among the earliest human implants and are increasingly utilized globally. Although exact numbers are unknown, more than 170,000 people in the United States had undergone cochlear implantation by 2015 \cite{nassiri_current_2022}. While cochlear implantation is more effective in prelinguistic children, its use in older populations is also rising \cite{rosenhall_otological_2011}. Beyond its social impact, CIs have become the most significant medical market \cite{sorkin_cochlear_2013}.     
A standrad CI comprises an external microphone, a speech processor unit, a transmitter, and flexible electrodes with 12-22 channels that stimulate the auditory nerve directly \cite{zeng_cochlear_2008} utilizing the tonotopic organization of the cochlea, where high-frequency signals are processed at the basal end and low-frequency signals are processed at the apical end of the cochlea \cite{von_bekesy_experiments_1960}. While traditional CIs have significantly improved the quality of life for individuals with severe to profound hearing loss, they present several challenges \cite{cohen_totally_2007}, such as weak adhesion for patients with thick hair, social stigmatization due to visible external components, frequent battery replacements, and issues like pressure sores and electric discharge. Fully/totally implantable cochlear implants (FICI/TICI) address these drawbacks by eliminating external parts, enhancing aesthetic appeal, user convenience, and overall device performance.  

However, the practical realization of FICI presents significant challenges. All implanted components must be fully biocompatible, compact, long-lasting, extremely low in power consumption, and immune to internal noise, such as heartbeats {or mechanical vibration from jaw movement or facial muscle contraction}. In contrast to the earlier concepts, where a special microphone was implanted in the skull, Envoy Medical has recently introduced a more progressive solution by detecting the movement of one of the ossicle bones, the incus, in the middle ear and converting the processed signal into the cochlea by the same multielectrode \cite{dornhoffer_initial_2023}. The main advantage of this solution is that it is {immune} to the internal body noises. Moreover, it does not sacrifice or bypass the middle ear organ, which functions well for most CI recipients, and it also utilizes the middle ear to probe the vibrations that were efficiently captured by the eardrum, taking advantage of natural filtering and amplifying characteristics of the outer and middle ear. 
Although the concept outlines a credible strategy for next-generation FICIs, the energy-efficient signal processing is still an issue, especially if a non-chargable implanted battery is used. {This persistent challenge underscores the need for novel signal processing strategies that address energy-efficiency at the architectural level, paving the way for practical, energy-efficient FICIs with extended battery life and potentially longer operational lifetime.} 

{The proposed architecture is built around three key components that together are designed to minimize power consumption and simplify auditory signal processing in fully implantable cochlear implants. These components include:} i) a Fast-Fourier Transformation (FFT)-free frequency-resolved cantilever array, ii) piezoelectric analog-signal generation, and iii) {a compact neuromorphic spiking element capable of directly generating and encoding action potential-like bipolar signals.} While the first two components have been detailed in our previous work~\cite{udvardi_spiral-shaped_2017}, this paper focuses on the third component and its compatibility with the preceding elements.


\begin{figure}[b!]
\centerline{\includegraphics[width=\columnwidth]{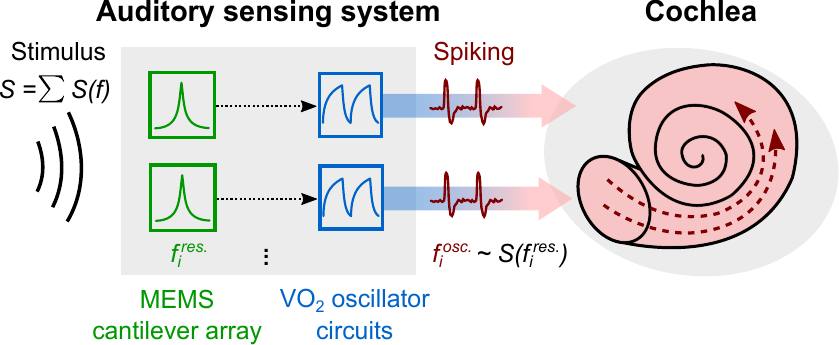}}
\caption{
Concept of our bio-inspired auditory sensing system. Vibroacoustic stimuli composed of several different frequency signals ($S=\sum S(f)$) are sensed by piezo-MEMS cantilevers, each having a well-defined resonance frequency, $f_i^{\rm res.}$, realizing frequency-selective sensing. The output signals of the cantilever array are carried to VO$_2$ memristor-based relaxation oscillator circuits ($i$ number of channels), which emit neural spikes proportional to the amplitude of the incoming stimulus in the $i^{\rm th}$ channel, $f_i^{\rm osc.}\sim S(f_i^{\rm res.})$. The auditory sensing system realizes rate-encoding in $i$ channels, creating spiking waveforms suitable for further processing by the nervous system, aimed at using in FICIs. } 
\label{fig:concept}
\end{figure}

{Building on these components, the auditory sensing system operates as illustrated in Fig.~\ref{fig:concept}:} Vibroacoustic stimuli with multiple frequency components ($S=\sum S(f)$) are sensed by a MEMS cantilever array, each element tuned to different resonance frequencies $f_i^{\rm res.}$, realizing frequency-selective sensing. {The resulting electric signals are routed to a set of compact neuromorphic oscillators}~\cite{CsabaGy2020} {which emit spiking signals proportional to the amplitude of the incoming stimulus in the $i^{\rm th}$ channel, $f_i^{\rm osc.}\sim S(f_i^{\rm res.})$. This realizes frequency-selective rate-encoding across $i$ channels and produces spike trains suitable for neuromorphic processing in FICIs.}

To implement this neuromorphic spiking element, we employ resistive switching memories, or memristors. These elements are anticipated to be crucial components in next-generation neuromorphic circuits due to their memory effect, small footprint, and low energy consumption~\cite{sebastian2020,wang2020,kumar2022,liu_artificial_2023,song2023,huang2024}. Non-volatile memristors show great promise for functioning as artificial synapses in spiking neural networks (SNNs)~\cite{zhou2022,boybat2018}. Volatile memristors, such as those constructed from VO$_2$ thin films, are excellent candidates for generating action potential signals, spiking/bursting patterns, or replicating several attributes of biological neurons, such as all-or-nothing firing, refractory period, spike frequency adaptation, and spike latency, thereby acting as artificial neurons~\cite{liu_artificial_2023,Yi2018,han_review_2022}.

In this paper, we take advantage of the capabilities of VO$_2$ memristors to demonstrate, for the first time, the direct conversion of analog MEMS signals into biomimetic {bipolar} signals similar to {biphasic signals} used in commercial implants~\cite{zeng_cochlear_2008,Navntoft2020}. The mechanical excitation applied within {a physiologically relevant nanometer-scale displacement range resembles the conditions inside the middle ear and is provided by a controlled source.} {Additionally, we demonstrate that the frequency of the generated spikes is proportional to the amplitude of the mechanical vibration, closely mimicking the human auditory system's temporal encoding of frequency-resolved signals}~\cite{rutherford2021}.

\section{Methods}

In this Section, we describe the measurement setup used for the characterization of MEMS cantilevers (Subsection~\ref{sec:MeasCanti}), provide details on our VO$_2$ nanogap memristors and relaxation oscillators (Subsection~\ref{subsec:memristor}), and validate our concept for an auditory sensing unit (i.e. one channel of the auditory sensing system shown in Fig.~\ref{fig:concept}, see Subsection~\ref{subsec:valiconcept}).



 
\subsection{Characterization of Piezoelectric MEMS Cantilevers}\label{sec:MeasCanti}

\begin{figure}[t!]
\centerline{\includegraphics[width=\columnwidth]{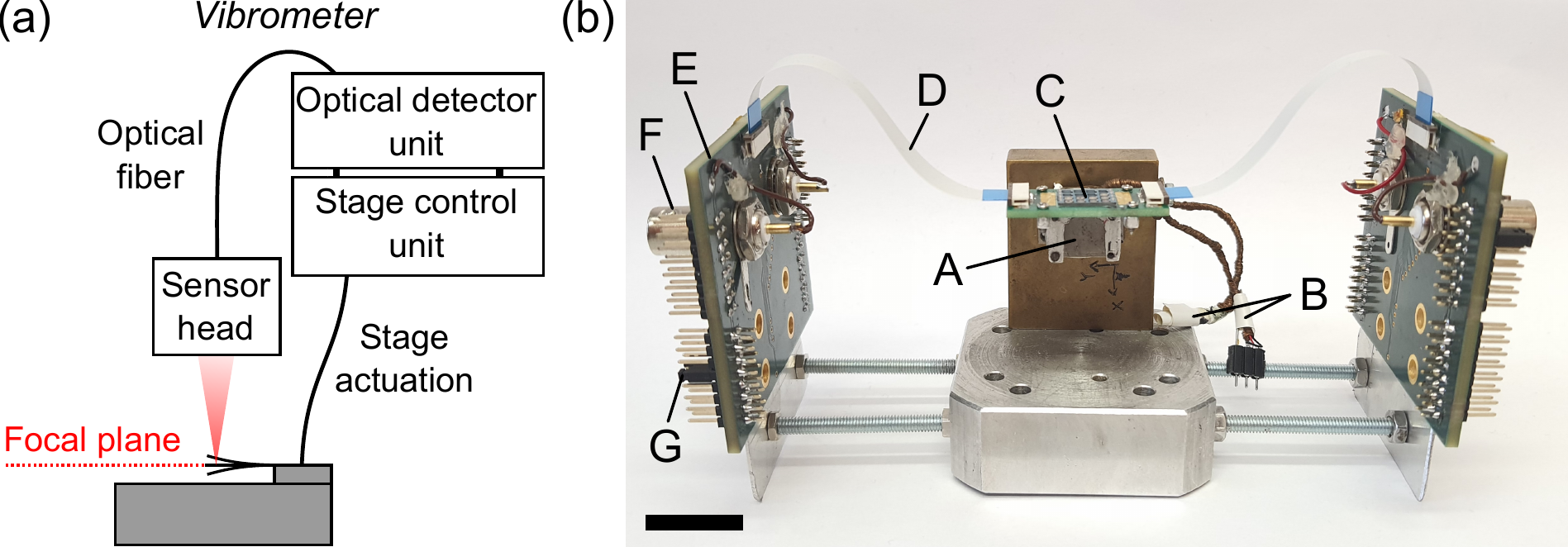}}
\caption{{Measurement setups for the characterization of piezo-MEMS cantilevers.} (a) Schematics of the interferometric measurement setup used for the mechanical characterisation of cantilevers. (b) Photo of the custom-built sample holder for the electromechanical test: (A) piezoelectric actuator, (B) shielded electrical connections to the piezoelectric actuator, (C) custom PCB holding the chip with cantilevers, (D) flexible ribbon cables to capture the cantilever's signal, (E) custom PCB for routing the connections to the cantilevers, (F) BNC connector, (G) jumper used for selecting the desired electrode of a cantilever. Scale bar: 2 cm.}
\label{fig:testing}
\end{figure}

The MEMS cantilevers used for our experiments {incorporate a piezoelectric ScAlN layer ~\cite{khanh_effect_2024} on a silicon-on-insulator platform.} The spiral-shaped design of the cantilevers was optimized for the targeted application previously~\cite{udvardi_spiral-shaped_2017}. To emulate conditions inside the middle ear, the cantilevers were put under $\sim$1~nm amplitude mechanical excitation, provided by the stage of a laser interferometric instrument (\emph{SmarAct PicoScale Scanning Vibrometer}), capable of recording the displacement of the cantilever. In this setup, illustrated in Fig.~\ref{fig:testing}a, the displacement is measured via the laser interferometric principle, and frequency spectra are recorded using the lock-in technique. Fig.~\ref{fig:testing}b shows the custom-built sample holder used for the electromechanical characterisation of the cantilevers. One important component of testing the presented concept for auditory sensing is providing biologically realistic, $\sim$10 nm excitation, i.e., in the range of the movement of bones in the middle ear. Mechanical excitation was provided by a piezo shear actuator (\emph{PI Ceramic P-142.03}), which has a sensitivity of 6~nm/V, and an axial resonant frequency of 120~kHz (far from the 200--700~Hz range under investigation). Nominal mechanical excitation amplitudes (stimuli) are calculated from the drive voltage amplitude of the piezoelectric actuator, based on the value of its 6~nm/V sensitivity. Electric signals of each cantilever are led through ribbon cables to custom PCBs on the two sides, where routing of each electrode is possible using two jumpers, leading the signal of a chosen cantilever to a pair of BNC connectors on the sides.

\subsection{Memristor-Based Oscillators}\label{subsec:memristor}

\begin{figure}[b!]
\centerline{\includegraphics[width=\columnwidth]{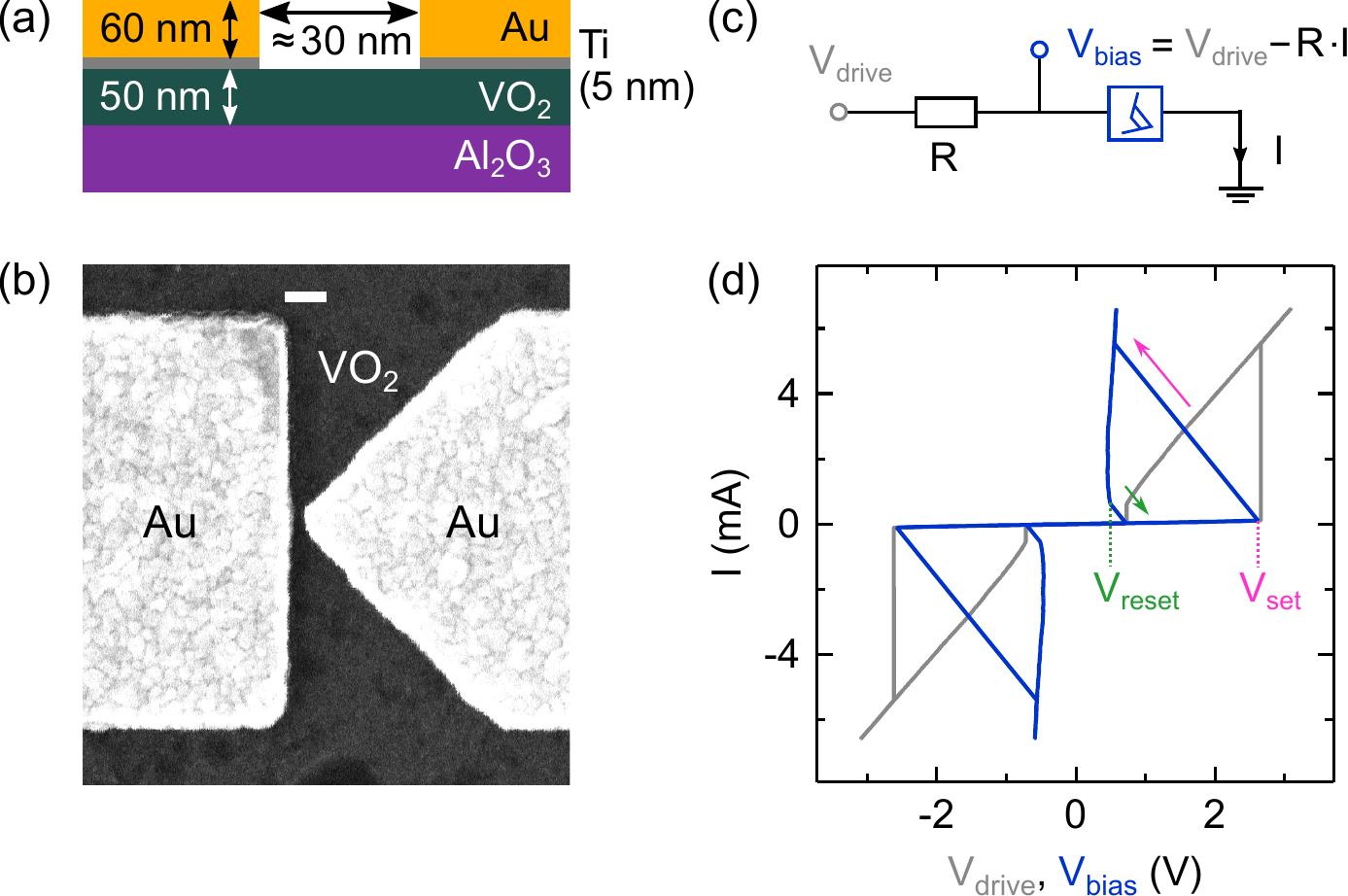}}
\caption{ 
Fabrication and characterization of VO$_2$ memristors. (a) Illustration of the planar memristor design showing layer thicknesses. (b) Scanning electron microscopy image of the nanogap region of a representative VO$_2$ device. Scale bar: 100~nm. (c) Circuit schematic for $I(V)$ characterization of the VO$_2$ samples. Drive voltage $V_{\rm drive}$ is applied to the memristor and a series resistor $R$, while current $I$ is monitored. Bias voltage is calculated as $V_{\rm bias} = V_{\rm drive}-R\cdot I$. (d) Typical $I(V)$ curves of a VO$_2$ device exhibiting volatile resistance switching, $R=380~\Omega$. Current is also displayed as a function of bias voltage (blue curve) and drive voltage (gray curve). The set process/IMT (reset process/MIT) is indicated by a pink (green) arrow on the $I(V_{\rm bias})$ curve. Note that these transitions are not resolved on the 12~kSa/s sampling rate of the measurement; there are no datapoints in the sections indicated with arrows.}
\label{fig:memristor}
\end{figure}

The composition of VO$_2$ memristors is illustrated in Fig.~\ref{fig:memristor}a.
For the fabrication of VO$_2$ memristors, 50 nm thick VO$_2$ layers were deposited on top of a sapphire (Al$_2$O$_3$) substrate, prepared via pulsed laser deposition. 
Layers were prepared according to the procedure reported in Ref.~\citenum{Kim2017}, without any buffer layers between the single crystal Al$_2$O$_3$ substrate and the VO$_2$. Stoichiometry and epitaxial growth of the VO$_2$ layer were confirmed by scanning transmission electron microscopy~--~electron energy loss spectroscopy (STEM EELS) and X-ray diffraction measurements (XRD)~\cite{Kim2014,Kim2015,Suess2017,Kim2017}.

On top of these epitaxial VO$_2$ layers, planar electrodes were deposited using electron-beam evaporation and lift-off techniques. The electrodes were patterned using standard electron-beam lithography (\emph{Raith 150}), and 50 nm Au was evaporated after a 5 nm Ti adhesive layer deposition. Ultra-small gaps between electrodes in the 30~--~60~nm range were formed with this method. The layout of the electrodes is depicted in the electromicrograph of Fig.~\ref{fig:memristor}b, which shows a V-shaped electrode facing a flat one. This arrangement ensures the focusing of the electric field into a small, nanometer-sized active region. The development of memristors with such a planar, nanogap layout was reported in Ref.~\citenum{Posa2023}.


Resistive switching characteristics of the VO$_2$ nanogap memristors were studied in the simple circuit arrangement depicted in Fig.~\ref{fig:memristor}c. Slowly ramped ($\sim1$~Hz) triangular voltage signals, $V_{\rm drive}$, are applied by a data acquisition card (National Instruments USB-6363) to an $R=380~\Omega$ series resistor and the memristor. The applied voltage $V_{\rm drive}$ and the current $I$ are measured using the same data acquisition card at 12~kSa/s sample rate, the latter is measured through a current amplifier (Femto DLPCA-200) in the $10^3$~V/A gain setting. Fig.~\ref{fig:memristor}d shows typical DC $I(V)$ characteristics of a VO$_2$ nanogap memristor, the grey curve depicts raw $I(V_{\rm drive})$ data, whereas the blue curve is the compensated $I(V_{\rm bias})$ dependence of the memristor, where the bias voltage $V_{\rm bias}=V_{\rm drive}-R\cdot I$ is the voltage drop on the memristor. Initially, at low $V_{\rm bias}$ levels, the memristor is in its insulating, high resistance state (HRS) until the voltage drop on it reaches the threshold for the onset of resistance switching (set process, see $ V_{\rm set}$ threshold in Fig.~\ref{fig:memristor}d). Once this threshold is reached, the memristor undergoes an insulator-to-metal transition (IMT, marked with a pink arrow), after which the memristor arrives in a conductive, metallic low resistance state (LRS). Upon decreasing the voltage, the memristor stays in its LRS until the $ V_{\rm reset}$ threshold is reached (here we define $V_{\rm reset}$ on the $I(V_{\rm bias})$ curve). Once $V_{\rm reset}$ is reached, the VO$_2$ memristor undergoes the reverse effect, metal-to-insulator transition ({MIT, marked with a green arrow}) is observed, and HRS is formed again.
The different resistances observed in these states (typically $R_{\rm LRS} \lessapprox 1~$k$\Omega$ and $R_{\rm HRS}\approx 30~$k$\Omega$~\cite{Pollner2025arXiv}) are the result of a Mott-type phase transition accompanied by structural rearrangement of the VO$_2$ material. 
In these nanogap memristors, the mutual effects of the high electric field, which stems from the ultra-small, $\approx30$~nm active region, and Joule heating are responsible for operation~\cite{Posa2023}.


\begin{figure}[t!]
\centerline{\includegraphics[width=0.75\columnwidth]{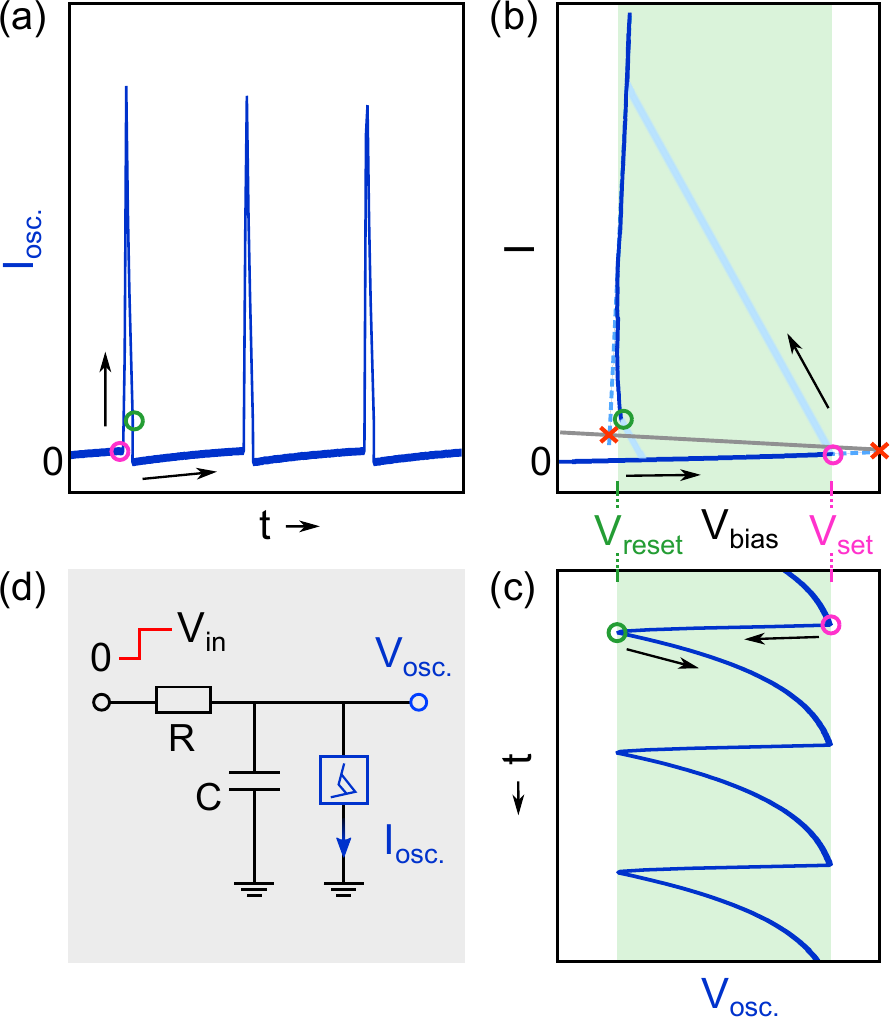}}
\caption{
Relaxation oscillator made of a VO$_2$ memristor. (a) Current waveform $I_{\rm osc.}(t)$ of an oscillator. (b) Magnified $I(V_{\rm bias})$ characteristics of a memristor (positive quadrant) shown with blue colors (transitions blurred with light blue). Grey line marks the possible equilibrium currents defined by the load line of the oscillator circuit according to the equation $(V_{\rm in} - V_{\rm bias} )/ R$. Equilibrium points defined by the continuation of the HRS and LRS are not reached, as marked by the red crosses. (c) Voltage waveform $V_{\rm osc.}(t)$ of an oscillator. The green shaded area and pink/green labels of $V_{\rm set}$/$V_{\rm reset}$ in panels (b-c) highlight the operation regime of the memristor during oscillation. Pink/green encircled points in panels (a-c) mark the points where set/reset transitions occur, and black arrows mark the direction. (d) Circuit schematics of the relaxation oscillator (output current/voltage and the memristor element are highlighted in blue).  }
\label{fig:VO2osc}
\end{figure}

A relaxation oscillator can be realized by connecting an $RC$-circuit in series with a VO$_2$ memristor (see Fig.~\ref{fig:VO2osc}).
The circuit shown in Fig.~\ref{fig:VO2osc}d exhibits oscillating current $I_{\rm osc.}$ (Fig.~\ref{fig:VO2osc}a) and voltage $V_{\rm osc.}$ (Fig.~\ref{fig:VO2osc}b) signals {while driven by} a constant input voltage of $V_{\rm in}$. 
During oscillation, the memristor switches back and forth between its LRS and HRS. The points where the set (reset) transition happens are illustrated by pink (green) circles in Fig.~\ref{fig:VO2osc}a-c. Note, that during oscillation, the whole hysteresis curve of the memristor is walked around, which means, that $V_{\rm osc.}$ varies between $V_{\rm set}$ and $V_{\rm reset}$ (see green shaded area and pink/green labels of $V_{\rm set}$/$V_{\rm reset}$ in Fig.~\ref{fig:VO2osc}b-c).

To illustrate the conditions required for stable oscillations, the positive quadrant of the memristor's $I(V_{\rm bias})$ curve in Fig.~\ref{fig:memristor}d is magnified in Fig.~\ref{fig:VO2osc}b. In this plot, set/reset transitions are blurred with a light blue color, since they are not resolved in DC measurements. With grey color, the load line of the oscillator circuit is shown, given by the equation $(V_{\rm in} - V_{\rm bias} )/ R$, which is a visualization of the possible equilibrium current flow in the circuit as a function of $V_{\rm bias}$.
The memristor would be in equilibrium if the load line intersected the HRS or LRS. Light blue dashed lines illustrate the continuation of these states, and red crosses mark their intersections with the load line; the current tends to reach these points in both states, but they are never reached since switching happens before that, which drives continuous oscillation.

The general role of circuit parameters affecting oscillatory behavior of a circuit depicted in the inset of Fig.~\ref{fig:VO2osc}d can be summarized as follows:
(i) The series resistance $R$ has to be chosen such that it ensures stable oscillation at the desired range of $V_{\rm in}$ input voltage levels (the load lines corresponding to $V_{\rm in}$ values do not cross HRS or LRS of the memristor); 
(ii) The parallel capacitance $C$ can be used to tune the oscillation frequency of the circuit to the desired timescale. The timescales of the circuit are primarily governed by the $RC$ time constants in the HRS and LRS of the memristor. Note that on the targeted, biologically relevant timescales in the $\sim$ms regime, the $\sim$ps$-$ns timescales of resistance switching in VO$_2$ memristors~\cite{schmid_picosecond_2024,Pollner2025arXiv} can be neglected, and switching can be considered as instantaneous.
(iii) The input voltage level also affects the oscillation frequency. While $R$ and $C$ are constant, this dependence can be {leveraged} to realize encoding of $V_{\rm in}$ levels in oscillation frequency.


\subsection{Concept Validation}\label{subsec:valiconcept}


\begin{figure}[b!]
\centerline{\includegraphics[width=\columnwidth]{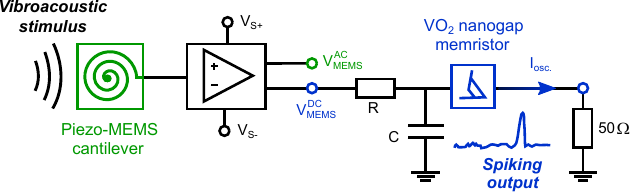}}
\caption{Circuit schematics of a single channel of the proposed auditory sensing system. A piezo-MEMS cantilever with a well-defined frequency resonance senses a sinusoidal vibroacoustic stimulus provided at the same frequency. As a result, an AC voltage signal is generated by the cantilever, which is amplified ($V_{\rm MEMS}^{\rm AC}$) and rectified ($V_{\rm MEMS}^{\rm DC}$) by a custom-built circuitry. The $V_{\rm MEMS}^{\rm DC}$ signal serves as an input DC voltage for the oscillator circuit, which emits current spikes at its output. These spikes are measured with the $50~\Omega$ input terminal of a digital oscilloscope. }
\label{fig:HW}
\end{figure}

For validating our concept of the auditory sensing system shown in Fig.~\ref{fig:concept}, we built and investigated the operation of a single channel according to the circuit depicted in Fig.~\ref{fig:HW}.
In our single-channel auditory sensing setup, a piezo-MEMS cantilever with a well-defined resonance frequency is driven by a sinusoidal vibroacoustic stimulus at the same frequency, generating an AC voltage. This signal is first amplified ($V_{\rm MEMS}^{\rm AC}$) and then rectified and smoothed to a DC level ($V_{\rm MEMS}^{\rm DC}$) by custom-made electronics designed for this purpose.
The resulting $V_{\rm MEMS}^{\rm DC}$ serves as an input to the VO$_2$ memristor-based oscillator circuit that produces output current spikes, which we record using the $50~\Omega$ input of a digital oscilloscope.

To verify rate encoding of the circuit due to biologically relevant, $\approx10$~nm amplitude mechanical stimuli, we designed a custom setup for the precise control of the stimulus amplitude (see more details in Subsection~\ref{sec:MeasCanti} and in Fig.~\ref{fig:testing}b). The operation of the circuit was also verified in less controlled experiments, where audio signals from a Bluetooth speaker were used as acoustic stimuli. Either a Picoscope 6404A instrument (electromechanical experiments) or a Tektronix DPO 3014 digital oscilloscope (audio experiments) was used to record the output signals of the circuit.


\section{Results}
\label{sec:results}

\subsection{Piezoelectric MEMS Cantilevers}

\begin{figure}[h!]
\centerline{\includegraphics[width=0.95\columnwidth]{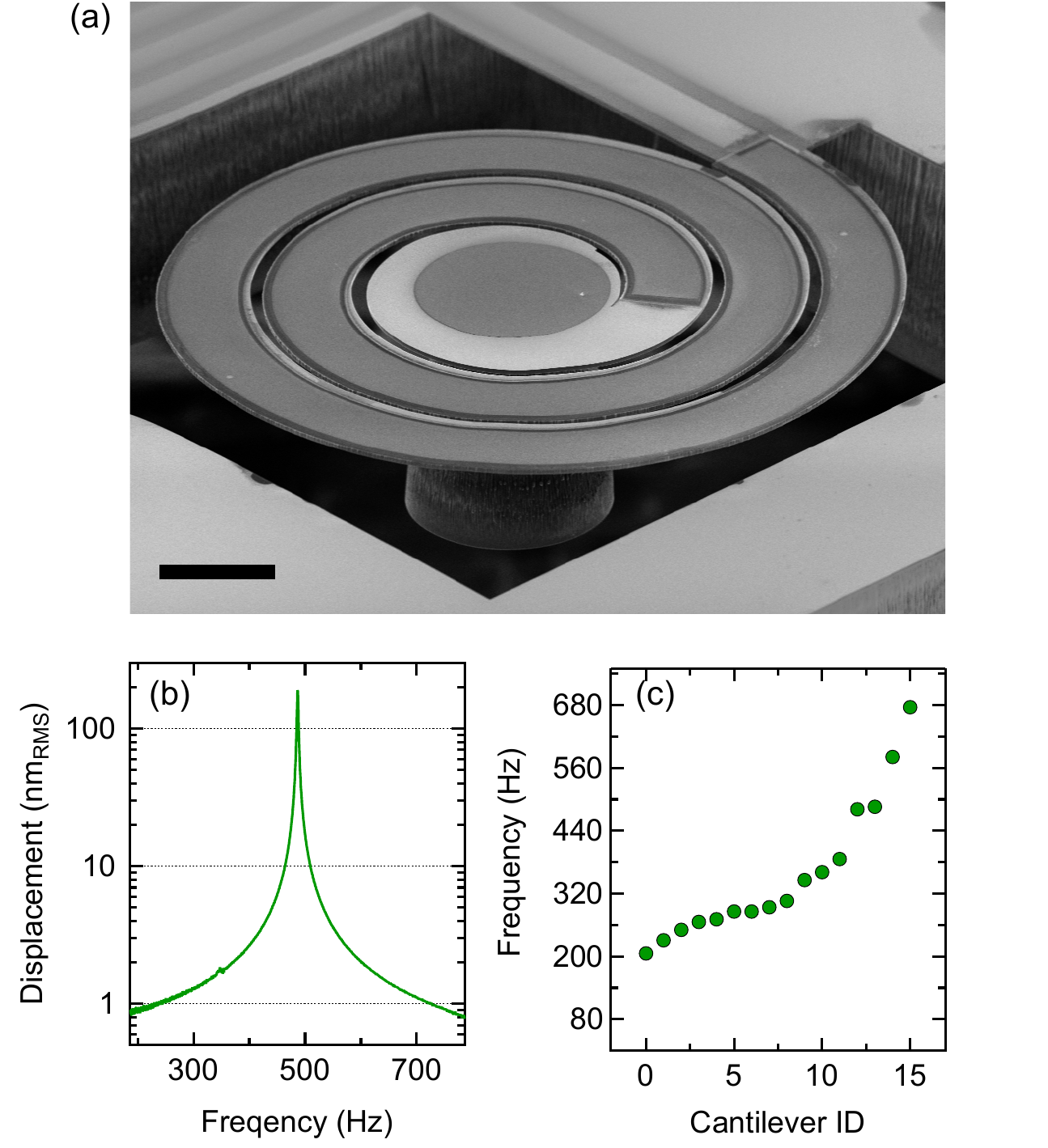}}
\caption{{Characterization of piezo-MEMS cantilevers.} (a) Tilt-view scanning electron micrograph showing the typical geometry of a piezoMEMS cantilever. Scale bar: 250~$\mu$m. (b) Displacement spectrum of a representative cantilever with $Q\approx200$ quality factor and $f_r = 486$~Hz resonance frequency. (e) Resonance frequencies of the studied 16 cantilevers covering the $200 -700$~Hz frequency domain.}
\label{fig:piezocantilever}
\end{figure}

A scanning electron micrograph of an exemplary spiral-shaped cantilever is shown in Fig.~\ref{fig:piezocantilever}a. The spiral is suspended at one side of a square cavity. The geometrical factors of the spirals, along with a seismic mass in the middle, determine their resonance frequencies~\cite{udvardi_spiral-shaped_2017}. 
The resonances are located at different frequencies, usually with $Q>200$ quality factor (see an exemplary resonance peak in Fig.~\ref{fig:piezocantilever}b recorded in a laser interferometric measurement). The resonance frequencies of the 16 cantilevers with slightly different geometries cover the desired frequency range of 200~--~700 Hz (see Fig.~\ref{fig:piezocantilever}c), where the fundamental frequencies of human speech are found. Conventional cochlear implants use $12-22$ channels in the same frequency domain~\cite{zeng_cochlear_2008}, being sufficient for speech recognition.

\subsection{Spike Generation and Rate-Encoding}

First, the response of the auditory sensing circuit was analyzed using controlled electromechanical stimuli through a piezoelectric actuator.
The cantilever chip was mounted to the sample holder shown in Fig.~\ref{fig:testing}b for these experiments.
The $V_{\rm MEMS}^{DC}$ electric signal of a cantilever was utilized to drive an oscillator circuit according to the circuit schematics in Fig.~\ref{fig:HW}. 

\begin{figure}[t!]
\centerline{\includegraphics[width=0.9\columnwidth]{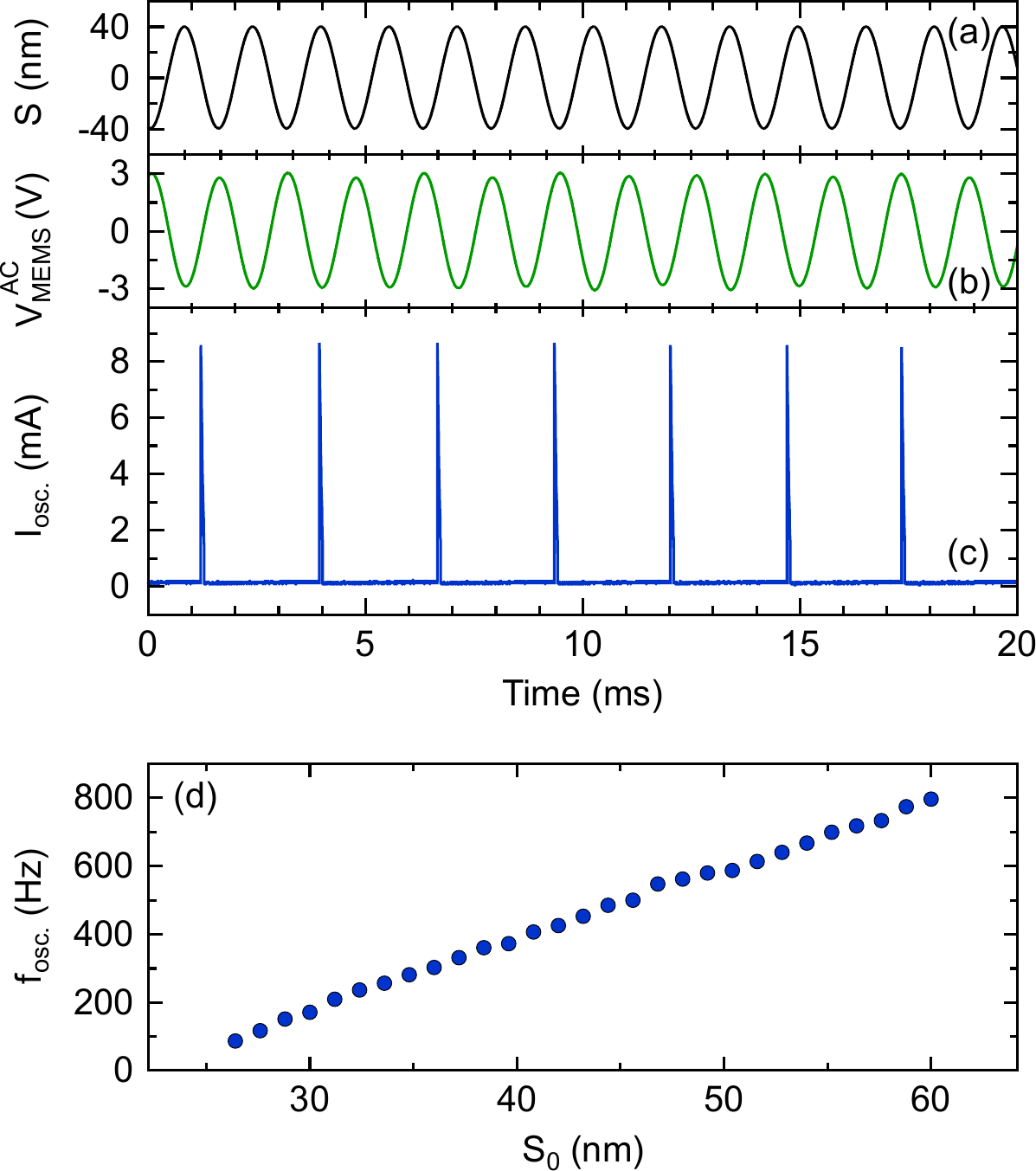}}
\caption{
Results of electromechanical experiments performed on a single-channel auditory sensing circuit, where stimulus to the cantilever is provided and controlled in the biorealistic displacement regime. (a) Waveform of a sinusoidal stimulus $S$, with constant frequency of 638 Hz. (b) Output signal $V_{\rm MEMS}^{\rm AC}$ of the cantilever, conditioned to $V_{\rm MEMS}^{\rm DC}=3.6~$V level, which drives the oscillator circuit. (c) Output current signal of the oscillator component exhibiting oscillation with a stable frequency. (d) Upon repeating experiments depicted in panels (a-c) with varying the stimulus amplitude $S_0$, we observed that the oscillation frequency $f_{osc.}$ depends linearly on $S_0$. Displacement values $S_0$ and $S$ are calculated from the voltage applied to the piezoelectric actuator and its 6~nm/V sensitivity. }
\label{fig:spike-generation}
\end{figure}

The typical result of an experiment is shown in Fig.~\ref{fig:spike-generation}a-c, where a MEMS cantilever is excited at its 638~Hz resonance frequency with an $S_0=40$~nm amplitude sinusoidal stimulus (see black waveform $S$). Due to this excitation, the cantilever emits a similar sinusoidal output (see amplified green signal, $V_{\rm MEMS}^{\rm AC}$),  which is conditioned to $V_{\rm MEMS}^{\rm DC}=3.6~$V level, driving the oscillator component of the circuit. 
As a result, a spiking current signal with stable oscillation frequency $f_{\rm osc.}$ appears at the output (see $I_{\rm osc.}$ with blue in Fig.~\ref{fig:spike-generation}c).
Note that $f_{\rm osc.}$ is entirely independent from the frequency at which the cantilever is excited, since the amplifying circuitry completely smoothes out AC components. The spiking frequency $f_{\rm osc.}$ is determined by (i) the values of $R$, $C$ elements in the oscillator circuit, (ii) the device characteristics of the VO$_2$ memristor (threshold voltages of $V_{\rm set}$, $V_{\rm reset}$ and resistance states $R_{\rm HRS}$, $R_{\rm HRS}$), and the (iii) input DC voltage level of the oscillator circuit ($V_{\rm MEMS}^{\rm DC}$ in this experiment). The oscillator circuit's parameters (i-ii) can be adjusted such that the resulting spiking frequencies correspond to biologically relevant spike rates. Since the input DC voltage level also affects spiking frequency and depends on the amplitude of the incoming stimulus, this chain of dependencies can be utilized for spike rate-encoding of the incoming vibration amplitude.

Overall, by appropriately setting $R$ and $C$ values in the oscillator component, the frequency range of possible oscillations can be set to the desired $\sim$1~kHz domain, where typical spiking rates of the auditory nerve are found~\cite{perez-gonzalez_adaptation_2014,bruce_phenomenological_2018}. {Furthermore, this input voltage dependency enables encoding the stimulus amplitude $S_0$ in the frequency of output spikes, as shown in Fig.~\ref{fig:spike-generation}d, where this was verified by varying the amplitude of sinusoidal stimuli. In this experiment, the piezo-MEMS cantilever was excited at its 637 Hz resonance frequency, with amplitudes in the 26~--~60~nm range. In a realistic setting where the cantilever would be mounted near the umbo, these amplitudes correspond to 82~--~92~dB sound pressure levels~\cite{Zurcher2010}. 
The parameters of the oscillator were chosen such that the resulting spiking frequencies are found in the 100~Hz~--~800~Hz range, which is common in biological conditions.}
The information encoding shown in Fig.~\ref{fig:spike-generation}d resembles the natural encoding of sound amplitude in the nervous system, with similar frequency--amplitude characteristics as the sigmoid-shaped functions observable at the cellular level~\cite{ziegler_bio-inspired_2023,westerman_conservation_1987}. {This rate-encoding behavior aligns with emerging fully implantable cochlear implant (FICI) concepts, which increasingly rely on neural-inspired temporal coding strategies for efficient auditory signal representation}~\cite{zak_model-based_2015}.

\subsection{Shape-Tuning of the Excitation Signal}

\begin{figure}[t!]
\centerline{\includegraphics[width=0.9\columnwidth]{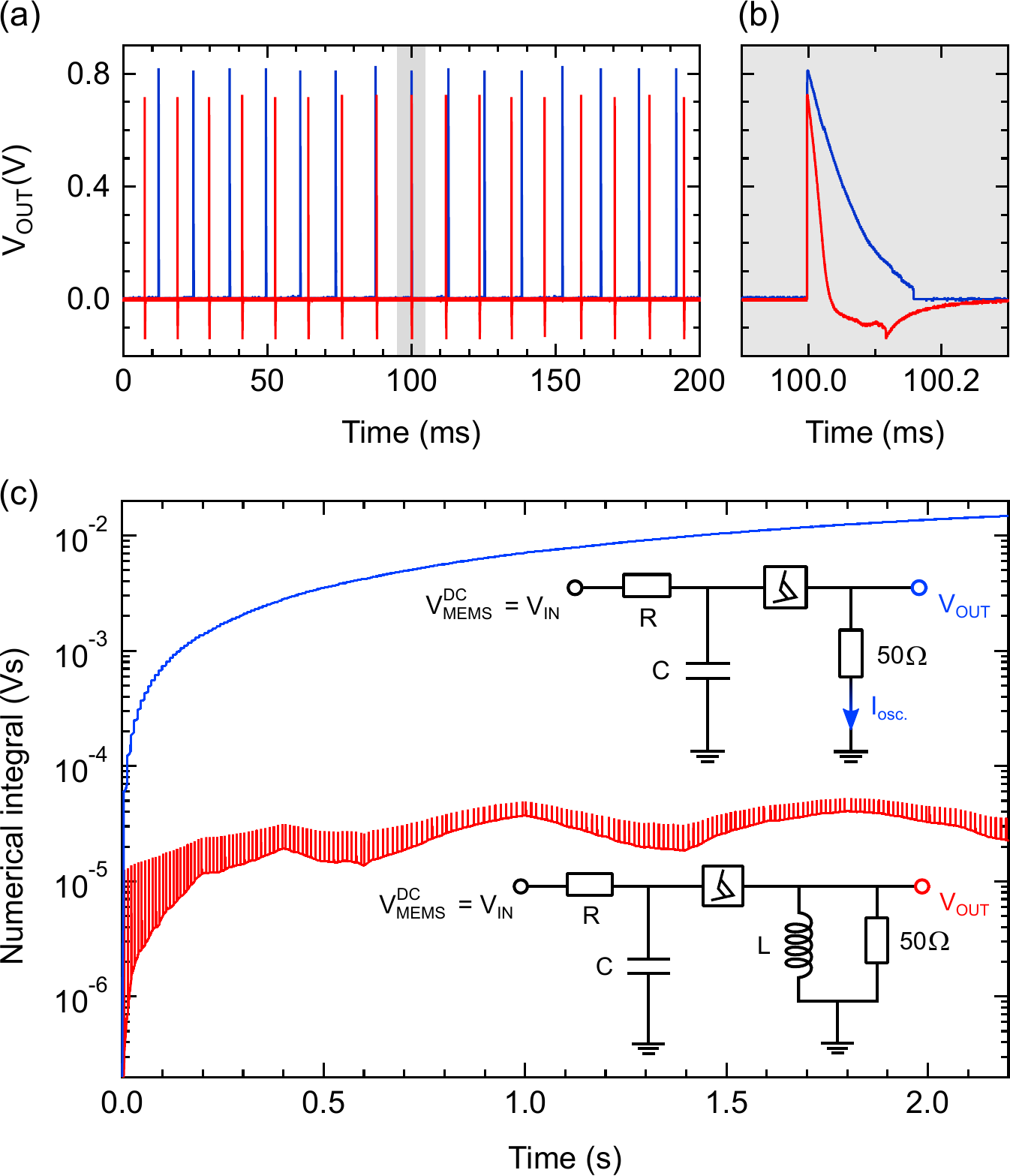}}
\caption{
Audio experiments with shape-tuning of the output waveform. (a) Comparison of output waveforms of the auditory sensing circuit built with a conventional oscillator (blue curve) and a modified oscillator (red curve). Only 200-ms-long sections of the 2.2-s-long measurements are shown; both waveforms are shifted such that a spike appears at 100 ms. (b) Plot is magnified on the time axis of the grey shaded area in panel (a) (the y-axis is the same as in panel (a)). (c) Comparison of numerical integrals calculated from the 2.2-s-long stimulation in case of the conventional oscillator (blue curve, see top inset for circuit schematics) and the modified oscillator (red curve, see bottom inset for circuit schematics) as the spiking component of the auditory sensing circuit. In the oscillator, $R=15~$k$\Omega$ and $C=611$~nF were kept constant, while an $L=2.2~$mH inductor was added to the modified circuit. }
\label{fig:signal-shaping}
\end{figure}

CI devices should satisfy an essential safety requirement: the spiking waveform carried to auditory nerves should be {bipolar (biphasic in CI terminology)}, yielding positive and negative sections as well. This {bipolar} waveform is needed to exclude the possibility of charge accumulation near the end of the implanted electrodes~\cite{zeng_cochlear_2008}. Current spikes at the output of a conventional VO$_2$ oscillator (see blue waveforms of Fig.~\ref{fig:signal-shaping}a-b and corresponding circuit schematics in the top inset of Fig.~\ref{fig:signal-shaping}c) are {unipolar (monophasic in CI terminology)}, as they only have positive sections. Therefore, additional circuit elements are needed to convert this waveform into a {bipolar} signal. This conversion can be realized by adding a parallel LR circuit to the output (see bottom inset of Fig.~\ref{fig:signal-shaping}c for schematics).

In our experiments, where we investigated shape-tuning, a less-controlled acoustic stimulus from a Bluetooth speaker held close to the cantilever array was used to provide stimuli. First, the operation of the circuit with a conventional oscillator component was verified (see blue waveforms in Fig.~\ref{fig:signal-shaping}a-b). 
Next, the appearance of a {bipolar} waveform was confirmed (see red signal in Fig.~\ref{fig:signal-shaping}a and magnified waveform in Fig.~\ref{fig:signal-shaping}b). The obtained waveform differs from the conventional rectangular pulses used in CI devices~\cite{zeng_cochlear_2008}; however, a study had shown that the usage of such ramped signals instead of rectangular ones (i) can induce similar electrically-evoked auditory brainstem response in mice, and (ii) could be more energy-efficient, having the potential to lower battery usage in CIs~\cite{Navntoft2020}. 

To demonstrate the difference between the cumulative effects of {unipolar} and {bipolar} signals, numerical integrals are calculated from 2~s-long experiments, shown in Fig.~\ref{fig:signal-shaping}c. As expected, the integral of the {unipolar} signal increases monotonically (blue curve). In contrast, the {bipolar} signal shows no definitive tendency; the fluctuation around a small, non-zero value is an instrumental artifact from a low-frequency drift superimposed on the measured signal.


\section{Conclusion and Outlook} 
\label{sec:conclusion}

We proposed a concept for an auditory sensing system designed for use in FICIs. The operation of a single channel was tested and verified in electromechanical experiments with controlled mechanical stimuli and also in audio experiments where stimuli were provided by a Bluetooth speaker. 
Our experiments demonstrate that the auditory sensing circuit is capable of emitting a spiking output due to biologically realistic excitation amplitudes in the $\sim10$~nm range and also capable of rate-encoding of the incoming stimulus amplitude, converting it to the $100\text{~Hz}-1\text{~kHz}$ output frequency domain.
In our experiments where audio signals were used as stimuli, we showed that {unipolar} current spikes can conveniently be transformed to a {bipolar} waveform as a prerequisite to application in cochlear implants.

The above findings can facilitate the application of the small-footprint, low-energy consumption auditory sensing circuit in fully implantable systems. The proposed circuit may also help to mitigate latencies of $\approx$10~ms typically experienced in commercial CI devices, i.e., the delay between the CI's electrical stimulation and the acoustic input, which has been shown to improve sound localization of patients~\cite{kortje2023}. In addition to the demonstrated rate-encoding approach, the proposed auditory circuit could also be adapted to support phase-locking through appropriate coupling of the MEMS output signal to the oscillator component. Since the auditory pathway inherently uses temporal information encoding principles like rate-encoding and phase-locking, applying these principles in cochlear implants can yield meaningful benefits~\cite{riss2014}.

{In summary, these results establish a low-power, biomimetic auditory sensing approach that may serve as a foundation for future fully implantable cochlear implant systems.}


\section{Acknowledgements}
This research was supported by the Ministry of Culture and Innovation and the National Research, Development and Innovation Office within the Quantum Information National Laboratory of Hungary (Grant No. 2022-2.1.1-NL-2022-00004), and the NKFI K143169, K143282, and TKP2021-NVA-03 grants. L.P. and T.N.T. acknowledge the support of the J\'{a}nos Bolyai Research Scholarship of the Hungarian Academy of Sciences.

\bibliographystyle{apsrev}
\bibliography{Auditory_Sensing.bib}

\end{document}